\documentclass[twoside,9pt]{article}
\usepackage{extsizes}
\usepackage[super,sort&compress,comma]{natbib} 
\usepackage[version=3]{mhchem}
\usepackage[left=1.5cm, right=1.5cm, top=1.785cm, bottom=2.0cm]{geometry}
\usepackage{balance}
\usepackage{mathptmx}
\usepackage{sectsty}
\usepackage{graphicx} 
\usepackage{lastpage}
\usepackage[format=plain,justification=justified,singlelinecheck=false,font={stretch=1.125,small,sf},labelfont=bf,labelsep=space]{caption}
\usepackage{float}
\usepackage{fancyhdr}
\usepackage{fnpos}
\usepackage[english]{babel}
\addto{\captionsenglish}{%
  
}
\usepackage{array}
\usepackage{droidsans}
\usepackage{charter}
\usepackage[T1]{fontenc}
\usepackage[usenames,dvipsnames]{xcolor}
\usepackage{setspace}
\usepackage[compact]{titlesec}
\usepackage{hyperref}
\usepackage{booktabs}
\usepackage{csvsimple}
\usepackage{longtable}
\usepackage{threeparttable}

\usepackage{xcolor}
\usepackage{soul}
\usepackage{siunitx}
\usepackage{braket}
\usepackage{amsmath}
\usepackage{dblfloatfix}

\newcommand{\doi}[1]{\href{http://dx.doi.org/#1}{\nolinkurl{#1}}}

\newcommand{\massU}{{\ensuremath{m/z}}}

\newcommand{\HHO}{{\ce{H2O}}}

\newcommand{\HH}{{\ce{H2}}}

\newcommand{\ie}{{i.\,e.}}
\newcommand{\eg}{{e.\,g.}}

\newcommand{\rcm}{\ensuremath{\text{cm}^{-1}}}

\newcommand{\nunup}{$2\nu_3^{2+}$}
\newcommand{\nunum}{$2\nu_3^{2-}$}
\newcommand{\op}{\ce{0+}}
\newcommand{\om}{\ce{0-}}

\newcommand{\HHHOp}{\ce{H3O+}}
\newcommand{\NN}{\ce{N2}}
\newcommand{\NHHH}{\ce{NH3}}

\definecolor{cream}{RGB}{222,217,201}

\begin{document}

\pagestyle{fancy}
\thispagestyle{plain}
\fancypagestyle{plain}{
\renewcommand{\headrulewidth}{0pt}
}

\makeFNbottom
\makeatletter
\renewcommand\LARGE{\@setfontsize\LARGE{15pt}{17}}
\renewcommand\Large{\@setfontsize\Large{12pt}{14}}
\renewcommand\large{\@setfontsize\large{10pt}{12}}
\renewcommand\footnotesize{\@setfontsize\footnotesize{7pt}{10}}
\makeatother

\renewcommand{\thefootnote}{\fnsymbol{footnote}}
\renewcommand\footnoterule{\vspace*{1pt}%
\color{cream}\hrule width 3.5in height 0.4pt \color{black}\vspace*{5pt}} 
\setcounter{secnumdepth}{5}

\makeatletter 
\renewcommand\@biblabel[1]{#1}            
\renewcommand\@makefntext[1]%
{\noindent\makebox[0pt][r]{\@thefnmark\,}#1}
\makeatother 
\renewcommand{\figurename}{\small{Fig.}~}
\sectionfont{\sffamily\Large}
\subsectionfont{\normalsize}
\subsubsectionfont{\bf}
\setstretch{1.125} 
\setlength{\skip\footins}{0.8cm}
\setlength{\footnotesep}{0.25cm}
\setlength{\jot}{10pt}
\titlespacing*{\section}{0pt}{4pt}{4pt}
\titlespacing*{\subsection}{0pt}{15pt}{1pt}

\fancyfoot{}
\fancyfoot[RO]{\footnotesize{\sffamily{1--\pageref{LastPage} ~\textbar  \hspace{2pt}\thepage}}}
\fancyfoot[LE]{\footnotesize{\sffamily{\thepage~\textbar\hspace{4.65cm} 1--\pageref{LastPage}}}}
\fancyhead{}
\renewcommand{\headrulewidth}{0pt} 
\renewcommand{\footrulewidth}{0pt}
\setlength{\arrayrulewidth}{1pt}
\setlength{\columnsep}{6.5mm}
\setlength\bibsep{1pt}

\twocolumn[
  \begin{@twocolumnfalse}
\par
\vspace{1em}
\sffamily
\begin{tabular}{m{4.5cm} p{13.5cm} }

 & \noindent\LARGE{\textbf{Near-infrared high resolution overtone spectroscopy of  
the hydronium ion \HHHOp: the \nunup\ and \nunum\ bands}} \\
\vspace{0.3cm} & \vspace{0.3cm} \\

 & \noindent\large{Chiara Schleif,\textit{$^{a}$} Hayley A. Bunn,\textit{$^{a}$} Miguel Jim\'enez-Redondo,\textit{$^{a}$}  
 Paola Caselli,\textit{$^{a}$} and Pavol Jusko$^{\ast}$\textit{$^{a}$}} \\

 & 
\noindent\normalsize{This work presents the rovibrational spectra of 
  the two strongest first overtone bands of the asymmetric stretching mode $\nu_3$ 
  (\nunup\ and \nunum) of the hydronium ion, \ce{H3O+}. 
  The measurements were performed in a temperature-variable cryogenic 22 pole ion trap 
  using leak-out spectroscopy (LOS), 
  covering an energy range of $6750-6950\;\rcm$. 
  The spectra were fit with a standard oblate symmetric top Hamiltonian and additional off-diagonal matrix elements accounting for $l$-doubling to obtain the spectroscopic constants and the band origins, which were determined to be $6845.610(14)\;\rcm$ for \nunup\ and $6878.393(13)\;\rcm$ for \nunum.} \\

\end{tabular}

 \end{@twocolumnfalse} \vspace{0.6cm}

  ]

\renewcommand*\rmdefault{bch}\normalfont\upshape
\rmfamily
\section*{}
\vspace{-1cm}


\footnotetext{\textit{$^{a}$~Max Planck Institute for Extraterrestrial Physics, Giessenbachstrasse 1, 85748 Garching, Germany. E-mail: pjusko@mpe.mpg.de}}


\section{Introduction}

\HHHOp\ is known to play a fundamental role in interstellar oxygen- and water chemistry.\cite{herbst1973,Phillips1992} 
It is produced by a well-known reaction chain starting from 
\ce{H3+} and \ce{O}, 
followed by subsequent reactions with \HH.\cite{herbst2001} 
\HHHOp\ itself does not react further with \HH\ and is eventually 
destroyed by dissociative recombination (DR)\cite{Jensen2000,herbst2001}
\begin{align}    
    \HHHOp\ +\ \ce{e-} &\rightarrow\ \HHO\ +\ \ce{H}~~~~~~~~~~\Delta H_r=-6.52\ \text{eV}\\
    &\rightarrow\ \ce{OH}\ +\ \HH ~~~~~~~~~~\Delta H_r=-5.89\ \text{eV}\\
    &\rightarrow\ \ce{OH}\ + \ce{H}\ + \ce{H}~~~~~\Delta H_r=-1.42\ \text{eV}\\
    &\rightarrow\ \ce{O}\ + \ce{H2}\ + \ce{H} ~~~~~~\Delta H_r=-1.48\ \text{eV}
\intertext{where (1) leads to the production of \ce{H2O} (enthalpies of the reaction channels are taken from the Active Thermochemical Tables (ATcT)\cite{Ruscic2005}).
The dissociation product \ce{OH} resulting from channels (2) and (3) can react further \textit{via} the reaction}
    \ce{OH}\ + \ce{O}\ &\rightarrow\ \ce{O2}\ +\ \ce{H}\ ~~~~~~~~~~~\Delta H_r=-0.705\ \text{eV}
\end{align}
to form \ce{O2}.\cite{Davidsson1990}
If \HHHOp\ is assumed to be the main precursor of both \ce{H2O} and \ce{O2}, 
its detection can be used to make order-of-magnitude predictions of the
abundances of the two components\cite{Phillips1992}, especially when the branching 
fractions and cross sections of the DR channels are known.\cite{Andersson1996,Jensen2000,Zhaunerchyk2009}

The DR of \HHHOp\ has been extensively studied in heavy-ion storage 
rings, e.g. with the CRYRING in Stockholm, Sweden or ASTRID in Aarhus, Denmark:
\citet{Neau2000} investigated the absolute cross sections of the DR of both
\HHHOp\ and \ce{D3O+}, \citet{Jensen2000} measured the cross sections of the different channels as a function of energy, and the branching ratios at E=0 for \HHHOp, \ce{HD2O+} and \ce{D3O+}. 
Other work focused on individual reaction channels, \citet{Andersson1996} for instance reported the first observation of the \HHO\ production path and its branching ratio, while
\citet{Zhaunerchyk2009} investigated the OH formation channel (2). 
It should be noted that the reported branching ratio of the \HHO\ production channel differs for all investigations, ranging from $17-18\;\%$\cite{Neau2000} up to $33\;\%$\cite{Andersson1996}.
Further work focusing on deuterated species like \ce{D3O+} has been done e.g. by \citet{Buhr2010} using the heavy-ion storage ring TSR in Heidelberg.

The first tentative detection of \HHHOp\ in the ISM was reported by 
\citet{Hollis1986} in 1986 towards the Orion-KL nebula \textit{via} a weak single-line detection. 
The observation was confirmed a few years later by \citet{Wootten1991} who discovered multiple 
rotational transitions towards OMC-1 and Sgr B2. In 2006, \citet{Tak2006} mapped \HHHOp\ in Sgr B2 using the APEX telescope.
The first detection in the far-infrared through metastable rotational transitions of hot \HHHOp\ was 
reported only much later in 2012 with Herschel in the same source, Sgr B2(N), by \citet{Lis2012}.\\ 
\begin{figure*}[b]
\centering
\vspace{2.5mm}
\begin{minipage}[t]{0.49\textwidth}
    \centering
    \includegraphics[width=0.98\linewidth]{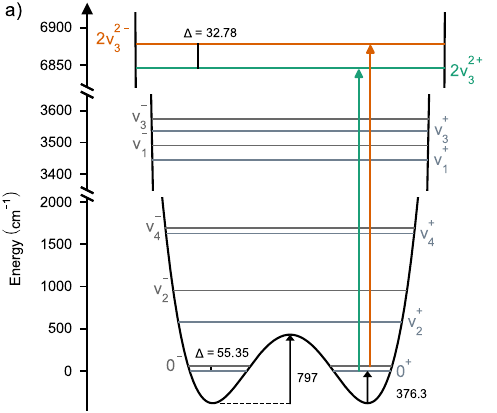}
\end{minipage}%
\begin{minipage}[t]{0.49\textwidth}
    \centering
    \includegraphics[width=0.98\linewidth]{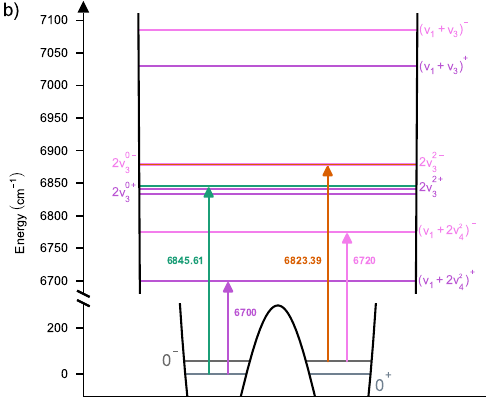}
\end{minipage}
\caption{Illustrative representations of the inversion potential energy function and vibrational levels of \HHHOp.
a) The potential function expanded by the energy levels of
\nunup\ and \nunum\ identified in this work. 
The energy levels of fundamentals and ground states were taken from \citet{Yu2009}, ground-state 
inversion splitting from \citet{Liu1985}, ZPE and inversion barrier from \citet{Sears1985}. 
The measured transitions are indicated by arrows in the colours of the respective final state.
b) An enlarged representation of the energy range relevant for this work, showing 
the relative locations of the overtone- and combination bands identified in the ExoMol data\cite{Yurchenko2020}. 
For comparison, the experimentally obtained energy levels
of the \nunup\ and \nunum\ band are included as well. 
Arrows indicate the allowed transitions from the ground states 
that are predicted to be most intense. 
The transition energies show that the \nunup\ and \nunum\ 
bands lie close together and well within the measured energy range, 
while the combination bands are expected rather further away.} 
\label{f_1}
\end{figure*}
In addition to the role of \HHHOp\ in the ISM, the hydronium ion was detected in 2009 by Cassini in the ionosphere of Enceladus, 
emerging from a cold water plume located in the south polar region\cite{Tokar2009}. 
Furthermore, laboratory investigations on exoplanetary atmospheres suggest \HHHOp\ to be a 
good candidate for future observations targeting the habitability of exoplanets:\cite{Bourgalais_2020,Bourgalais_2021}
Bourgalais and his team 
concluded that \HHHOp\ could be expected in 
sub-Neptunes\cite{Bourgalais_2020} and might be one of the most abundant ions in Titan-like
atmospheres containing traces of \HHO.\cite{Bourgalais_2021} \\
\HHHOp\ is a pyramidal symmetric top molecule and isoelectronic to \NHHH. Just like ammonia, 
it can be described by the molecular symmetry group D$_{3h}$ due to its relatively low inversion
barrier in the potential energy function\cite{Bunker} which is visualised in Figure \ref{f_1} a). 
This low barrier of 797 \rcm\cite{Sears1985} results in the observed inversion-splitting of all vibrational levels into $+$/$-$ 
states.\\ 
The first experimental high-resolution infrared spectrum of \HHHOp\ was measured by 
\citet{Begemann1983} in 1983, including a preliminary analysis of the doubly degenerate asymmetric stretching mode $\nu_3$. 
Two years later, the group published a more extensive analysis of the same degenerate vibration\cite{Begemann1985}, 
reporting strong Coriolis and Fermi interactions between the various modes. \citet{Begemann1985} observed a large standard deviation of the fit when 
neglecting those perturbations and concluded that a simultaneous analysis of multiple bands is needed to improve the large residuals.
In parallel, \citet{Liu1985v2} reported the observation of both inversion 
components of the $\nu_2$ band and the rotational and quartic centrifugal distortion constants resulting from a least-squares fit. Based on these measurements they determined the inversion 
splitting of the ground states to be $55.3462(55)\;\rcm$\cite{Liu1985}, which is much higher than the value observed for \NHHH ($0.8\;\rcm$\cite{Good1946}). In 1989, \citet{Verhoeve1989} 
measured the full inversion spectrum between \op\ and \om, observing interactions between states with $\Delta K = \pm 3n$, which they included in the fit to obtain improved spectroscopic constants for the ground states.
\citet{Tang1999} investigated the stretching modes $\nu_1$ and $\nu_3$ as well as the ground 
states \op\ and \om\ again after a previous study\cite{Ho1991} in great detail in 1999,
observing Coriolis interactions and \textit{l}-type doubling between the stretching modes. They reported rotational and quartic centrifugal distortion constants for all investigated modes and additionally Coriolis coupling and l-doubling constants for $\nu_3$. 
The first investigations on the doubly degenerate bending mode, $\nu_4$, were published by \citet{Gruebele1987} in 1987, where they reported rotational and quartic/sextic centrifugal distortion constants. The group observed significant Coriolis interactions with the $\nu_2$ and $2\nu_2$ levels, but did not consider those perturbations in the fit.
More recent work by \citet{Yu2009} from 2009 combined previous data of the four fundamental modes
and the ground state inversion transition with new measurements of the latter, taking into account 
the observed strong perturbations by incorporating the Coriolis interaction terms between $\nu_1$ and $\nu_3$ 
and the $\Delta K = \pm 3$ interactions
between the ground states \op\ and \om\ in their analysis. Inclusion of these coupling terms allowed for an assignment of 200 additional high $\textit{J}$-transitions 
and improved rotational, up to octic centrifugal distortion and Coriolis coupling constants for the ground states and all fundamentals.\\
While the fundamental modes of \HHHOp\ are known in great detail, no rotationally resolved experimental work 
has been done on its overtones. Only recently, \citet{Huang2024} published near-infrared
spectroscopy measurements of $\HHHOp\;\cdot\;$X$_{n}$ clusters (X=Ar, \NN, CO; $n=1-3$) covering the energy region where the first overtones of the stretching modes are expected. 
They were able to identify several overtone- and combination bands of the ion-tag cluster, but without rotational resolution. 
Within the ExoMol project\cite{Tennyson2012}, \citet{Yurchenko2020} published a thorough theoretical rovibrational 
molecular line list for \HHHOp, covering energies up to 10,000 \rcm\ and temperatures up to 1500 K. 
The line list is based on an \textit{ab initio} dipole moment surface, an empirical potential energy surface 
which was globally fitted to the available rovibrational energies of the ground 
and fundamental vibrational states and variational nuclear motion calculations done with the program TROVE\cite{YURCHENKO2007}. \\
In contrast to the hydronium ion, the overtones of \NHHH\ have been investigated in much greater detail: 
As an example, \citet{SUNG2012} examined the energy region between $6300-7000\;\rcm$ and assigned over 1000 transitions to seven overtone and combination bands, mentioning 
the high amount of perturbations observed and expected between the states. 
In 2020, \citet{FURTENBACHER2020} used the MARVEL database to further expand and improve the accuracy of known rovibrational energy levels in the energy range of up to 7500 \rcm.

The aim of this paper is to gain more knowledge of the first overtones of \HHHOp:
This work presents the first high resolution measurements of the strong overtone bands \nunup\ and \nunum, two of the first overtones of the doubly degenerate, asymmetric stretching mode $\nu_3$. 

\section{Experimental details}
The measurements of the near-infrared spectrum of \HHHOp\ were performed in the CCIT 22 pole cryogenic ion trap, which has been previously 
described in detail\cite{Jusko2024}.
The \HHHOp\ ions ($19\;\massU$) were produced in a Storage Ion Source by electron bombardment of a precursor
gas mixture consisting of the vapour pressure of \HHO\ at room temperature (approx. $0.023\;\text{atm}$) topped up with \HH\ to $1\;\text{atm}$. The ions extracted from the source were then
mass-selected with a quadrupole filter and focused into the trap, where they were decelerated 
and cooled down with short Helium pulses from a piezo valve, obtaining temperatures close to the overall trap temperature. 
A Helium cryostat connected to the trap chamber allows for efficient cooling down to $4\;\text{K}$ and enables, in combination with a heating element, flexible temperature regulation.
In this experiment, the trap temperature was kept constant at $125\;\text{K}$ and $20\;\text{K}$.
An Agilent 8164B option 142 light system was used 
with the the module covering $1370-1495\;\text{nm}$, a line width < 100 kHz and wavelength-dependent
CW power between $1-8\;\text{mW}$. 
The laser was calibrated using the Bristol 671A wavemeter. To excite a maximum number of ions,
the laser was aligned in a triple-pass through the trap cell using a setup of optical mirrors\cite{Redondo2024a}. 
As the number of ions in the trap is not sufficient to enable the use of direct absorption spectroscopy,
action spectroscopy needs to be done instead. 
In the beginning, it was intended to apply a laser-induced reaction (LIR) scheme 
using {\ce{C2H4}} and {\ce{C3H8}} as reaction partners. However, this turned out not to be feasible
due to the rapid agglomeration of both neutral species with {\HHHOp}.
Consequently, leak-out spectroscopy (LOS)\cite{Schmid2022} was chosen to record the overtone spectrum, 
where \NN\ and \HH\ were used as neutral buffer gases. 
The latter was chosen specifically for the measurements at $20\;\text{K}$, below the freezing point of \NN.

The general concept of LOS is described in detail in \citet{Schmid2022}.
In brief, the neutral buffer gas \NN\ (\HH) is used as collision partner
to induce a transfer of the internal energy into kinetic energy, 
which enables the ions to overcome a potential barrier at the exit and ``leak out'' of the trap. 
The potential barrier is tuned in such a way that this process is enhanced when the photon wavelength 
is on resonance with a rovibrational transition of the \HHHOp\ ion, leading to an increase in the observed ion
signal.
The trapping cycle was set to 3 s (LOS acquisition time around 2.2 s) and a continuous inflow 
of \NN\ (\HH) into the trap chamber throughout the whole measurement cycle
resulted in a number density of neutral molecules of around $10^{11}\;\text{cm}^{-3}$. 
Although \HH\ is only $\sim 10\;\%$ of the mass of \HHHOp, there was no impact on the performance of LOS in comparison to the use of \NN. 
This further supports the premise, that coupling between vibrational transitions of the two molecules is more important for vibration-transition energy transfer, than the pure momentum conservation due to mass.

An energy range of $1481-1438\;\text{nm}$ 
(corresponding to $6750-6950\;\rcm$) was scanned in $0.0002-0.0004\;\text{nm}$ steps 
to obtain the spectra reported in this work. 
Further scanning towards higher and lower energies beyond this energy range within the capacity of the laser did not reveal more transitions.
The ExoMol data available for \HHHOp\ \cite{Yurchenko2020} that fell into
the operation range of the laser module was used to estimate a starting point for the measurements. 
Every line identified in the scans was re-measured multiple times and the resulting average then used for the line profile and position fit.

\begin{figure}[h]
\centering
    \includegraphics[width=0.99\linewidth]{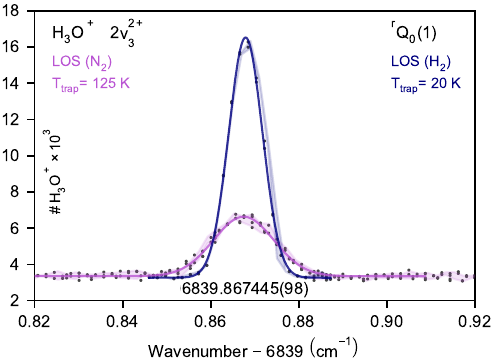}
    \caption{Comparative representation of a line profile measured at both $125\;\text{K}$ (purple)
    and $20\;\text{K}$ (navy). The black number at the bottom of the plot depicts the fitted line position in \rcm,
    which got assigned to the transition $^\text{r}\text{Q}_0(1)$ of the 2$\nu_3^{2+}$ band. 
    The line profiles at $20$ and $125\;\text{K}$ were measured with an average of $48$ and 
    $235$ thousand primary ions in the trap, respectively.
    Transparent areas in the background represent the std. error around the binned signal 
    (0.002$\;\rcm$ bin width).
    }\label{f_2}
\end{figure}

For the measurements at $20\;\text{K}$, each transition detected at $125\;\text{K}$ was first tested for visibility and 
then re-scanned to obtain the new line width and intensity at the lower temperature.
An exemplary comparison of a line profile visible at both operation temperatures is given in 
Figure \ref{f_2}. A change in line width and intensity is clearly evident, although the latter
is additionally impacted by the total number of ions in the trap, which can vary significantly between measurements.

\section{Results and Discussion}
In total, 85 transitions were observed at $125\;\text{K}$, of which 36 are also visible at $20\;\text{K}$. 
The resulting spectra are shown in Figure \ref{f_3} and a complete line list is given in Table S1 
in the Supplementary Information (SI). 
The spectrum at $125\;\text{K}$ was first compared to a refined version of the ExoMol data that contained 
additional information to the transitions, \ie, initial and final state and the respective TROVE\cite{YURCHENKO2007} quantum numbers and symmetry 
labels $\Gamma_{vib}$ and $\Gamma_{rot}$ assigned.\cite{Yurchenko2020}
The quantum numbers given in the data file can be correlated to the normal modes $\nu_1 - \nu_4$ and the inversion symmetry of the respective state. 
Together with a comparison to the known overtone- and combination bands of \NHHH\cite{FURTENBACHER2020},
this enables an approximate separation 
of the calculated transitions into the bands that overlap within the energy region.
Furthermore, expected values for the angular momentum quantum numbers $l_3$ and $l_4$ of degenerate states can be identified. 
An overview of the theoretically predicted bands separated from the ExoMol dataset is given in Table \ref{t2}, including an 
estimation of their band origin where possible. 
A schematic visualisation of the relative positions of the involved states is given in Figure \ref{f_1} in panel b). 
Based on the intensity distribution among the bands seen in the ExoMol data file, the strongest bands in the observed energy range 
appear to be 2$\nu_3^{2+}$ and 2$\nu_3^{2-}$, with some transitions of the combination bands 
$(\nu_1$+2$\nu_4)^{\pm}$ and $(\nu_3$+2$\nu_4)^{\pm}$ having comparable intensities as the weaker 
transitions of the 2$\nu_3^{2\pm}$ bands. 
However, it is important to note that the refined version of the dataset only corresponds to a temperature of $300\;\text{K}$, \ie,
the intensity distribution will differ at $125\;\text{K}$.
Additional factors in the experiment relating to the collision process can lead to variations in the intensities, thereby line assignments based exclusively on intensity are unreliable and should be avoided.
\begin{table}[h!]
    \begin{threeparttable}
        \centering
        \caption{Selection of theoretically predicted overtone- and combination bands of \HHHOp\ in the energy range measured in this work, based on ExoMol.\cite{Yurchenko2020}}
        \begin{tabular*}{0.48\textwidth}{@{\extracolsep{\fill}}lcc}
            \toprule
            \textbf{Band}\tnote{a} & \textbf{Symmetry} & \textbf{Estim. Origin $\left(\rcm\right)$} \\
            \midrule
            2$\nu_3^{2-}$          & E" \tnote{b}     & $\sim\,6879$ \\
            2$\nu_3^{2+}$          & E' \tnote{b}     & $\sim\,6841$ \\
            
            2$\nu_3^{0-}$          & A$_2$" \tnote{b} & $\sim\,6878$ \\
            2$\nu_3^{0+}$          & A$_1$' \tnote{b} & $\sim\,6833$ \\
            
            $(\nu_1 + \nu_3)^-$    & E" \tnote{b}     & $\sim\,7085$ \\
            $(\nu_1 + \nu_3)^+$    & E' \tnote{b}     & $\sim\,7030$ \\
            
            $(\nu_1$+2$\nu_4)^-$   & E" \tnote{b}     & $\sim\,6775$ \\
            $(\nu_1$+2$\nu_4)^+$   & E' \tnote{b}     & $\sim\,6700$ \\
            
            $(\nu_3$+2$\nu_4^2)^-$ & E", A$_2$", A$_1$" \tnote{c,d} & / \\
            $(\nu_3$+2$\nu_4^2)^+$ & E', A$_1$' \tnote{c,d} & / \\
            $(\nu_3$+2$\nu_4^0)^{\pm}$ & E \tnote{e} & / \\
            \bottomrule
        \end{tabular*}
        \label{t2}
        \begin{tablenotes}
            \item[a] Inversion symmetry and vibrational angular momentum quantum numbers $l_3, l_4$ got
            assigned based on band symmetry as known for the fundamentals and expected for the overtones 
            based on MARVEL database entries of \NHHH\cite{FURTENBACHER2020}.
            \item[b] Based on MARVEL database entries of \NHHH\cite{FURTENBACHER2020} and 
            confirmed by the ExoMol data.
            \item[c] Based on MARVEL database entries of \NHHH\cite{FURTENBACHER2020} but 
            no direct confirmation with the ExoMol data possible.
            \item[d] Expected splitting based on MARVEL database entries of \NHHH\cite{FURTENBACHER2020}. 
            \item[e] Based on \citet{SUNG2012} but no direct confirmation with the ExoMol data possible.
        \end{tablenotes}
    \end{threeparttable}
\end{table}
The best match was obtained when comparing the measurements to the separated, theoretically predicted \nunup\ and \nunum\ bands, which raised the assumption that the majority of the measured transitions belong to either 
one of these two. While the pattern of the predicted subbranches was consistent with the 
measurements, single line positions did not agree (the comparison is visualised in Figure S1 in the SI). A detailed analysis of the measured spectrum 
was, therefore, done using the program PGOPHER\cite{WESTERN2017221}, a versatile tool to fit and simulate molecular spectra. 
The rotational hamiltonian used for symmetric tops\cite{WESTERN2017221} (only including the spectroscopic constants that could be fit in this work) is
\begin{equation}
\begin{split}
    H_{rot} =\ &Origin\ +\ BN(N+1)\ +\ (C-B)K^2\\
    &+\ (-2C\zeta+\eta_JN(N+1)\ +\ \eta_KK^2)lK\\
    &-\ D_JN^2(N+1)^2\ -\ D_{JK}N(N+1)K^2\ -\ D_KK^4\\ &+\ H_JN^3(N+1)^3\ +\ H_KK^6, 
\end{split}
\end{equation}
where B and C are rotational constants, D and H quartic and sextic distortion constants, and $\zeta$ and $\eta$ Coriolis coupling constants. 
The off-diagonal matrix element included to account for the $l$-doubling in this case is 
\begin{equation}
\begin{split}
    &\braket{N, K+2, l+1|H|N, K, l-1} = \\
    &\frac{1}{2}q_+\big[(N(N+1)-K(K-1))(N(N+1)-K(K+1))\big]^{\frac{1}{2}} 
\end{split}
\end{equation}
with the $l$-doubling constant $q_+$.
Similar to \NHHH, only states with the symmetries A$_2$', A$_2$", E' and E" are allowed for \HHHOp. The statistical weights of states with A$_1$' and A$_1$" symmetry are zero, as can be derived by nuclear spin statistics described by \citet{Bunker}. In this context, the only difference between \NHHH\ and \HHHOp\ is the nuclear spin quantum number \textit{I} of the central atom (=1 for $^{14}$N and 0 for $^{16}$O), which results in the representation $\Gamma_{spin}$ being a factor of 3 higher for \NHHH.
The forbidden and allowed state symmetries as well as the relations between the statistical weights are, therefore, identical for both molecules.\\
The molecular constants of the ground states \op\ and \om\ 
obtained by \citet{Yu2009} (listed in Table S2 in the SI) were used as initial states to predict the transitions, also considering the $\Delta K = \pm 3$ interaction between the two states
by including the off-diagonal matrix elements 
\begin{align}
    &\braket{N,\ K\pm3,\ l|J_zJ_{\pm}^3|N,\ K,\ l} = \nonumber\\ 
    &\sqrt{2}\alpha\times(2K\pm3)\sqrt{
    \begin{aligned}
        &(N(N+1)+(-K\mp1)(K\pm2))\times \\
        &(N(N+1)+(-K\mp2)(K\pm3))\times \\
        &(N(N+1)-K(K\pm1))
    \end{aligned}
    }
\end{align}
with $\sqrt{2}\alpha=8.506\;\text{MHz}\equiv0.0002837\;\rcm$.
Linear extrapolations of the spectroscopic constants for the fundamentals 
$\nu_3^+$ and $\nu_3^-$ obtained by the same work\cite{Yu2009} (listed in Table S2 in the SI) were used as an initial prediction for the simulation of the overtones. 

The two overtone bands \nunup\ and \nunum\ are known from \NHHH\ to be perpendicular (E'/E" symmetry) 
and therefore require a conservation of state symmetry\cite{SUNG2012}. \nunup\ is, therefore, excited from the 
\op\ and \nunum\ from the \om\ ground state. 
Referring to the predictions from ExoMol\cite{Yurchenko2020}, it was evident that the two bands are 
overlapping heavily, which made a reliable assignment difficult for the majority of the transitions. 
Consequently, the spectrum was re-measured at $20\;\text{K}$ to depopulate the \om\ ground state almost completely and thus minimize transitions owing to the \nunum\ band.
The resulting spectrum is seen in Figure \ref{f_3} a). 
The \nunup\ band was fit using the remaining visible transitions and later refined 
by stepwise assignment of increasing $J$ transitions from the $125\;\text{K}$ spectrum that showed residuals below $0.1\;\rcm$ (in this work, the term ``residuals'' refers to the difference between experimental line position obtained from the line profile fitting and simulated line position). 
Some transitions could not be fit although they were assigned reliably. They stayed with a higher deviation and were, consequently, excluded from the fit and marked with * in Table S1 in the SI. 
The remaining transitions in the $125\;\text{K}$ spectrum were assigned to the \nunum\ band, 
which was then fit in a similar procedure.
The transitions predicted at $35\;\text{K}$, resulting from the fit at $125\;\text{K}$, well reproduced the observed transitions visible at $20\;\text{K}$ as visualised in Figure \ref{f_3} a). This provided further confidence in the fit of the two bands.
\begin{figure}[t!]
\centering
    \includegraphics[width=0.99\linewidth]{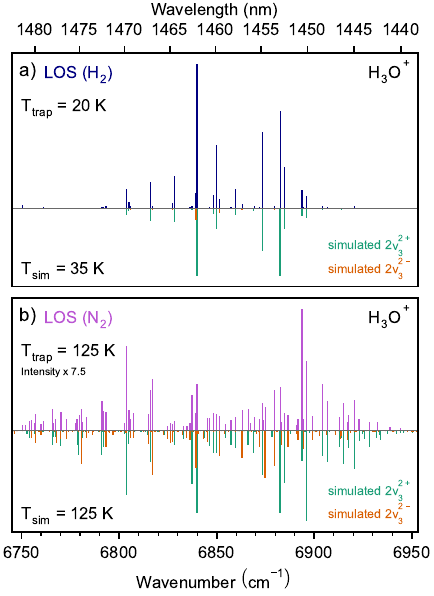}
    \caption{Comparison of the measured and simulated overtone spectrum of \HHHOp\ at $20\;\text{K}$ (panel a) 
    and $125\;\text{K}$ (panel b, scaled by factor 7.5 relative to panel a) for comparability).
    The measurements cover an energy range of $6750-6950\;\rcm$. The simulations were obtained by fitting 
    a Hamiltonian to 68 of the 85 transitions, where 39 transitions got assigned to the \nunup\ band (simulation in teal), the other 29 transitions
    to the \nunum\ band (simulation in orange). The resulting parameters are listed in Table~\ref{t1}. For further details see text.}
  \label{f_3}
\end{figure}
Even though the temperature of the trap chamber was measured to be $20\;\text{K}$, the ions are expected 
to be slightly warmer, which is confirmed by comparing the simulated spectra at $20\;\text{K}$ and $35\;\text{K}$.
Around $125\;\text{K}$, the slight temperature difference between ions and trap chamber has a negligible effect on the simulated spectrum.
Consequently, the simulation set to the experimental temperature reproduces the spectrum well, as can be seen in Figure \ref{f_3} b).\\
In total, 77 of the 85 measured transitions could be assigned to the two overtone bands, of which 68 were 
included for the fitting procedures (39 transitions for \nunup, 29 for \nunum). The exclusion of 8 transitions was due 
to large remaining shifts as explained above. One of the transitions could be assigned certainly only at $20\;\text{K}$, 
while several assignments became possible at $125\;\text{K}$. This transition was, therefore, not included in the fit either.
All assignments are listed in Table S1 in the SI.
The 8 transitions that remain unassigned are expected to either belong to another band, \eg, 
($\nu_1$ + 2$\nu_4^2$)$^\pm$, or are perturbed so heavily that no certain assignment is possible. \\
\citet{Ho1991} reported for the bands $\nu_3^\pm\leftarrow 0^\pm$ that $^PP$-transitions of $J''=4,\ 5,\ 7,\ 8\ (\nu_3^+\leftarrow 0^+)$ and 
$J''=4,\ 5,\ 8,\ 12\ (\nu_3^-\leftarrow 0^-)$ showed large residuals and had to be excluded from the fit. 
The group was able to prove that the deviations are expected to be caused by perturbations, but
its source could not be determined. \citet{Tang1999} extended the assignment of the $\nu_3^\pm\leftarrow 0^\pm$ bands to higher
$J,\ K$ transitions: They observed an increase in the residuals proportional to $J-K$, specifically for $^PP$-transitions with $K'\geq7$.

Of the 77 transitions assigned in this work, 18 turn out to be $^PP$-transitions, and only 4 of the 8 heavily shifted transitions. 
A pattern within the residuals of this type of transition could, therefore, not be determined reliably. 
Instead, the analysis focuses on patterns in the upper state quantum numbers $J'\;\text{and}\;K'$. 

The majority of the transitions got assigned to $J'= 2, 3, 4$ and $K'=0, 1, 2, 3$. 10 transitions got assigned to $J'=5$ and 2 to $J'=6$. 
No transitions with $J'\;\text{and}\;K'> 7$ have been observed. Only one transition got assigned to $J',K'=7$. 
In regard of $|J-K|$, 36 transitions show $|J-K|=0$, 21 $|J-K|=1$, 13 $|J-K|=3$ and only 2 $|J-K|=4$, both belonging to $J'=4$. A comparison of the change in residuals with increasing $|J-K|$ for $J'=4$ did not reveal any proportional trend as it was seen for higher $J'$ for $\nu_3^\pm$ by \citet{Tang1999}. 
In general, for the 8 heavy shifted transitions, no pattern could be identified within the involved $J'\;\text{and}\;K'$ states. 
The only apparent aspect is that solely $J',K'=3,4,5$ seem to be sufficiently affected. 
When looking at transitions with residuals~$>0.01\;\rcm$, it becomes evident that especially transitions with $J'=3$ show larger deviations, while $J'<3$ are less affected. 
Higher $J'$ transitions are seen at both higher and lower residuals, so no clear pattern can be drawn. 

A trend in the deviations is expected to become clearer when transitions with higher $J'\;\text{and}\;K'$ are incorporated. 
This assumption is supported by \citet{Tang1999}, where clear trends are observed especially for $K'$ $\geq7$. 
No such transitions were measured within this work, as the line intensities 
are expected to be significantly weaker and therefore not visible with this experimental setup.

The spectroscopic constants obtained from the fit of \nunup\ and \nunum\ are listed in Table \ref{t1}, 
including 1$\sigma$ uncertainty in the last two significant digits in parentheses.
Apart from $H_J$ (and $H_K$ for \nunup), no higher distortion constants could be fit with reasonable errors, which is why they were put to zero, likely due to the lack of transitions to higher $J'$ and $K'$ states.
\begin{table}[h!]
    \centering
    \caption{Spectroscopic constants obtained for the overtone bands \nunup\ and \nunum\ 
    of \ce{H3O+}, values given in $\rcm$ with 1$\sigma$ in parentheses.}
    \begin{tabular*}{0.48\textwidth}{@{\extracolsep{\fill}}lS[table-format=4.9]S[table-format=4.9]}
        \toprule
        \textbf{Parameter} & $\mathbf{2\nu_3^{2+}}$ & $\mathbf{2\nu_3^{2-}}$ \\
        \midrule
        $Origin$ & 6845.610(14) & 6878.393(13) \\
        $B$ & 10.8905(43) & 10.7716(25) \\
        $C$ & 6.1091(36) & 6.1851(47) \\
        $D_J$ & 0.00173(33) & 0.00024(11) \\
        $D_{JK}$ & -0.00318(23) & -0.00171(32) \\
        $D_K$ & 0.00265(39) & 0.00296(40) \\
        $\zeta$ & 0.03784(47) & 0.02813(70) \\
        $\eta_J$ & -0.00303(53) & 0.0064(11) \\
        $\eta_K$ & 0.00524(50) & -0.0028(11) \\
        $q_+$ & 0.2099(18) & -0.2239(16) \\
        $H_J\ \times10^5$ & 3.24(76) & 0.0 \\
        $H_K\ \times10^5$ & -3.71(94) & 5.33(60) \\
        \bottomrule
    \end{tabular*}
    \label{t1}
\end{table}
Despite excluding the largest deviations from the fit, many line positions remain shifted with respect
to the simulations, as can be seen in Table S1 in the SI. The best fit of the \nunup\ band has
an average error of $0.0265\;\rcm$, the \nunum\ band $0.0147\;\rcm$. 
The main cause of this is expected to be some strong perturbations caused by the bands predicted nearby, 
visualised in Figure \ref{f_1}. In the past, strong Coriolis and Fermi interactions with neighbouring bands were not only seen for the fundamentals,
described e.g. by \citet{Tang1999} and \citet{Yu2009}, but also for the fundamental, combination and overtone bands 
of \NHHH\ (described e.g. in \citet{SUNG2012}). Therefore, similar interactions are also expected for the overtone and combination bands of \HHHOp.
Since the two bands identified in this work are separated by only $32.78\;\rcm$ and have both E'/E" symmetry, 
they are prone to perturb each other through Fermi interaction.
Attempts at implementing this interaction and also perturbation by unobserved states were not successful 
and did not obtain a meaningful improvement of the residuals. An effective fit to unperturbed transitions is, therefore, reported here. 
It is expected that inclusion of higher $J'$ transitions will assist in correctly modelling the interactions responsible for the deviations. 
In fact, \citet{Yu2009} reported that considering the strong Coriolis interactions between the fundamental stretching modes $\nu_1$ and $\nu_3$ 
and inclusion of 200 more high-$J$ transitions enabled an improvement of the molecular parameters and frequency predictions of further high-$J$ transitions. 
Furthermore, gaining information about the unobserved nearby states is required to treat the interaction in the fits. 
The combination bands $(\nu_1$+2$\nu_4)^\pm$ are, according to the ExoMol predictions, expected to be
the next most intense bands nearby the measured energy region and could play an important role in the observed perturbations. 
To further investigate these assumptions, more measurements are needed, covering the energy range of the bands 
predicted nearby and sensitive to significantly weaker transitions. For this, a different laser light source would be needed.

\section{Conclusions}
In this work, the two degenerate overtone bands \nunup\ and \nunum\ have been identified and fit using
a standard oblate symmetric top Hamiltonian to obtain two sets of spectroscopic constants, including rotational, centrifugal distortion, 
Coriolis coupling and l-type doubling constants.
Apart from $H_J$ and $H_K$, no higher order centrifugal distortion constants could be obtained.
Although interaction between the identified bands and severe perturbations by unobserved nearby bands were observed, no interaction parameters could be determined and therefore only the effective fits to unperturbed transitions are reported. 
The identification of further combination- and overtone bands, \eg\ $(\nu_1$+2$\nu_4)^\pm$, is expected 
to facilitate the implementation of perturbation terms, together with additional measurements of higher $J$-state transitions for \nunup\ and \nunum. Such measurements are also expected to enable the determination of higher centrifugal distortion constants.\\
In comparison with the high-resolution measurements obtained in this work, the calculations from 
ExoMol\cite{Yurchenko2020} are in agreement regarding subbranch structure and separation, but disagree at the level of rotational resolution and therefore prevent a direct line assignment. 
The theoretical band origins deviate in the order of \rcm, with the prediction of {\nunum} being close 
to the experimental value, while {\nunup} shows a slightly higher deviation. 
Running a cross-correlation between the transitions separated from the ExoMol data and from the experiment 
showed that a shift to the band origin alone cannot successfully align all rotational transitions at once
(see Figure S2 in the SI).
It is, therefore, expected that there are further factors involved to cause the disagreement between predictions and experiment. A fit to the experimental transitions as done in this work is, consequently, essential to provide improved rotational line positions.\\
As the calculations from ExoMol refer to $300\;\text{K}$, while the measurements were done at $125\;\text{K}$, 
the simulated spectra were additionally extended to $300\;\text{K}$ (using the obtained spectroscopic constants listed in Table {\ref{t1}}) in order to match the ExoMol data temperature. 
The comparison is shown in Figure S3 in the SI. 
The agreement of several transitions is again improved when shifting the band origin of the ExoMol calculations, but,
similar to the comparison with the experimental data, no alignment of all transitions can be obtained. 
It is, therefore, to be expected that the fit to the experimental transitions obtained in this work already
improves the predictions of higher $J$ and $K$ transitions, despite neglecting the observed perturbations.

Besides supporting the improvement of theoretical calculations, the spectroscopic data obtained in this work is expected to have a wide range of possible applications, 
as cheap telecommunication equipment covering the E and S-band can now be used for \emph{in situ} monitoring in emission or absorption.
This directly allows for further investigation of the dissociative recombination of \HHHOp\ in plasma discharge, the use of spectroscopic identification instead of mass spectrometry in breath analysis\cite{Spanel2019}, 
or its detection 
in hot environments like exoplanetary atmospheres.

\subsection*{Conflicts of interest}

There are no conflicts to declare.

\subsection*{Data availability statement}

The data (raw measurements, post-procesing scripts, line lists, and, pgopher file)
that support the findings of this study are openly available at 
{\small \url{https://doi.org/10.5281/zenodo.14645564}}.

\section*{Acknowledgements}

This work was supported by the Max Planck Society.
The authors gratefully acknowledge the work of the electrical and mechanical workshops and engineering 
departments of the Max Planck Institute for Extraterrestrial Physics.
We thank Prof. Stephan Schlemmer (Univ. zu K\"{o}ln) for lending of the Agilent laser system.

\bibliography{lit} 
\bibliographystyle{rsc} 


\end{document}